\documentclass{emulateapj}
\usepackage{apjfonts}

\newcommand{\ang}{\AA}
\newcommand{\civ}{{\sc C iv}}
\newcommand{\feii}{Fe {\sc ii}}
\newcommand{\hb}{${\rm H}\beta$}
\newcommand{\hbi}{${\rm H}\beta_{\rm IC}$}
\newcommand{\hbb}{${\rm H}\beta_{\rm BC}$}
\newcommand{\hbn}{${\rm H}\beta_{\rm NC}$}
\newcommand{\hbvb}{${\rm H}\beta_{\rm VBC}$}
\newcommand{\kms}{$\rm km~s^{-1}$}
\newcommand{\ks}{$\chi^2$}

\newcommand{\myemail}{chenhu@bao.ac.cn}

\newcommand{\oiii}{{\sc [O iii]}}

\slugcomment{Accepted for publication in ApJ (Letters).}

\shorttitle{Intermediate-line Region in Quasars}
\shortauthors{Hu et al.}

\begin{document}

\title{\hb\ Profiles in Quasars:
Evidence for an Intermediate-line Region}

\author{Chen Hu\altaffilmark{1,2,3}, Jian-Min Wang\altaffilmark{2,4},
Luis C. Ho\altaffilmark{5}, Yan-Mei Chen\altaffilmark{2,3}, 
Wei-Hao Bian\altaffilmark{2,6} and Sui-Jian Xue\altaffilmark{1}}

\altaffiltext{1}{National Astronomical Observatories of China,
Chinese Academy of Sciences, Beijing 100012, China; \myemail}
                                                                                
\altaffiltext{2}{Key Laboratory for Particle Astrophysics,
Institute of High Energy Physics, Chinese Academy of Sciences,
Beijing 100049, China.}
                                                                                
\altaffiltext{3}{Graduate University of the Chinese Academy of Sciences,
Beijing 100049, China.}

\altaffiltext{4}{Theoretical Physics Center for Science Facilities (TPCSF), 
Chinese Academy of Sciences, China.}

\altaffiltext{5}{The Observatories of the Carnegie Institution of Washington,
813 Santa Barbara Street, Pasadena, CA 91101, USA.}

\altaffiltext{6}{Department of Physics and Institute of Theoretical Physics,
Nanjing Normal University, Nanjing 210097, China.}

\begin{abstract}
We report on a systematic investigation of the \hb\ and \feii\ emission lines 
in a sample of 568 quasars within $z < 0.8$ selected from the Sloan Digital 
Sky Survey.  The conventional broad \hb\ emission line can be decomposed into 
two components---one with intermediate velocity width and another with very 
broad width. The velocity shift and equivalent width of the intermediate-width 
component do not correlate with those of the very broad component of \hb, but 
its velocity shift and width do resemble \feii.  Moreover, the width of the very broad component is 
roughly 2.5 times that of the intermediate-width component. These 
characteristics 
strongly suggest the existence of an intermediate-line region, whose 
kinematics seem to be dominated by infall, located at the outer portion of the 
broad-line region.  
\end{abstract}

\keywords{galaxies: nuclei --- (galaxies:) quasars: emission lines ---
(galaxies:) quasars: general --- galaxies: Seyfert --- line: profiles}

\section{Introduction}

The geometry and kinematics of the broad-line region (BLR) in active galactic
nuclei (AGNs) have been studied for about three decades but the details are
far from well understood.  It is widely accepted that the BLR is stratified:
high-ionization lines originate from small radii and low-ionization lines
arise further out \citep{collin88}. This stratification picture is
supported by the results of reverberation mapping, which show that lines of 
different ionization have different lags (e.g., \citealt{peterson99}).  
The dependence of the systemic velocities of the emission lines on ionization
\citep[e.g.,][]{gaskell82,sulentic00a,richards02,shang07} suggests that the 
BLR may originate from a wind and a disk \citep[e.g.,][and references 
therein]{leighly04a,leighly04b}.  However, the profiles of the broad emission 
lines often contain multiple velocity components, suggesting that the
structure of the BLR may be more complex than can be described by a 
simple stratification or wind $+$ disk model.

It has been known that the broad \hb\ line profiles are generally not well
described by a single Gaussian. Two Gaussians \citep[e.g.,][]{netzer07} or a
Gauss-Hermite function \citep[e.g.,][]{salviander07} are often used.
Additionally, the profiles show great diversity from object to object.  Sources
with narrower \hb\ lines tend to have stronger line wings, while those 
with broader \hb\ lines are dominated by the line core 
\citep[e.g.,][]{sulentic02}. Some sources have asymmetric and shifted \hb\ 
profiles, suggesting that the BLR has a structure more complex than a single 
virialized component. The \hb\ profile of OQ 208, for example, has an 
additional redshifted \hb\ component of intermediate width that closely
resembles the kinematics of \feii\ \citep{marziani93}. Additional evidence 
that the BLR contains two or more kinematically distinct components comes 
from differential variability between the line core and wing
\citep[e.g.,][]{ferland90,peterson00}.  The existence of an
intermediate-line region and a very broad-line region for \hb\ emission was
suggested by some previous studies \citep[e.g.,][]{brotherton96,sulentic00b}.

In a recent spectral decomposition of a large sample of quasars selected from 
the Sloan Digital Sky Survey (SDSS), Hu et al. (2008, hereinafter Paper I) 
find that the majority of quasars show \feii\ emission that is both redshifted 
and narrower than \hb.  Moreover, the magnitude of the \feii\ redshift 
correlates inversely with the Eddington ratio.  These characteristics suggest 
that \feii\ originates from an exterior portion of the BLR, whose dynamics may 
be dominated by infall.  These findings offer fresh insights into the 
structure of the BLR. 

In light of the trends associated with \feii\ emission, we expect that a 
portion of the \hb-emitting gas may be related to an inflowing component too.
In this Letter, we systematically study the \hb\ profiles of SDSS quasars to
try to answer two questions: do all quasars have an intermediate-width \hb\ 
component similar to OQ 208, and, if so, is this component also associate with 
the \feii\ emission?  We show that the conventional broad \hb\ line actually 
consists of two kinematically linked components, one of which originates
from the same region that emits \feii.

\section{Sample and Data Analysis}

Our sample is selected from the SDSS Fifth Data Release (\citealt{adelman07}) 
quasar catalog \citep{schneider07}.  We choose objects with redshifts $z<0.8$ 
to ensure that \oiii\ $\lambda$5007 lies within the SDSS spectral coverage. We 
also require a signal-to-noise ratio (S/N) $>$ 10 in the restframe wavelength 
range 4430--5550 \ang\ (covering \hb, \oiii, and the most prominent features 
of optical \feii\ emission) and that no more than 1/3 of the pixels are masked 
by the
SDSS pipeline in this region. 7601 quasars satisfy these criteria.  The 
spectral analysis, whose details are described in Paper I, involves fitting a 
continuum model in a set of windows devoid of strong emission lines that 
consists of (1) a single power law, (2) Balmer continuum supplemented with 
high-order Balmer emission lines, and (3) a pseudo-continuum due to blended
\feii\ emission.  The full width at half maximum (FWHM) and shift of \feii\ are
measured from the continuum decomposition. After subtracting the continuum 
model, the \hb\ line is decomposed into narrow (\hbn) and broad (\hbb) 
components.  \hbn\ is forced to have the same profile as \oiii, a shift of up 
to 600 \kms\ relative to \oiii, and an intensity constrained to lie between 
1/20 and 1/3 of that of \oiii. \hbb\ is modeled using a Gauss-Hermite function 
\citep{vandermarel93}, whose best fit yields FWHM(\hbb).
The rest frame is defined by the peak of the \oiii\ $\lambda$5007 (see Paper
I for more details).

\begin{figure}
  \includegraphics[angle=-90,width=0.47\textwidth]{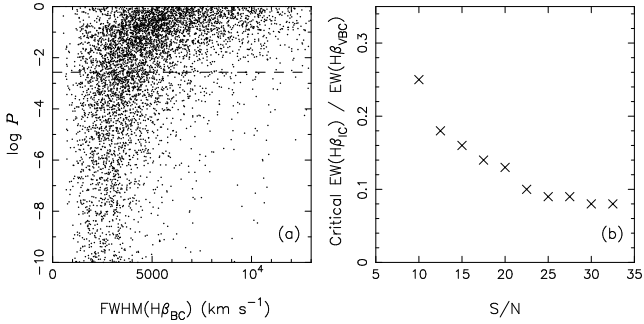}
\caption{\footnotesize ({\it a}) Variation of \hb\ profile with line width. 
$P$ denotes the probability that the double-Gaussian model cannot improve 
the fit.  A high value of $P$ means that the \hb\ line can be adequately 
described by a single Gaussian, whereas a low value of $P$ indicates that 
two Gaussians are needed.  The dashed line is $1-P = 99.73 \%$. The sources 
below this line need two Gaussians to fit their broad \hb\ lines. ({\it b}) 
The critical EW ratio of \hbi/\hbvb\ for requiring the double-Gaussian model
as a function of signal-to-noise ratio.} 
\label{fig-hbwfp}
\end{figure}

Although the Gauss-Hermite function provides a good and convenient mathematical 
description of the \hbb\ profile, for the purpose of this
Letter---investigating whether there is an intermediate-width \hb\ component
associated with \feii\ emission---we use the following simpler method to
fit \hbb. First, we use only one Gaussian to model the broad \hb\ line
(hereinafter the single-Gaussian model).  Then we fit broad \hb\ with two 
Gaussians (hereinafter the double-Gaussian model), an intermediate-width 
component (\hbi) and a very broad component (\hbvb).  In both models, \hbn\ is 
fitted in the same way as in Paper I.  First, we compare the \ks\ of the
double-Gaussian model and the Gauss-Hermite. There are 630 sources with 
\ks\ 20\% larger in the double-Gaussian model than in the Gauss-Hermite model
\citep{hao05}; this means that the double-Gaussian fit is still inadequate for 
these sources\footnote{From a statistical view for the whole sample, however,
this does not mean that the double-Gaussian model is worse than the 
Gauss-Hermite model. There are 510 sources with \ks\ 20\% smaller for the 
double-Gaussian model than for the Gauss-Hermite model, and for most sources 
both models are equally good.}.
Then, for the sources that can be well fitted by the
double-Gaussian model, we compare the reduced \ks\ of the double-Gaussian
and single-Gaussian model, and use the F-test \citep[Chapter 12.1]{lupton93}
to calculate how significantly the double-Gaussian model improves the fit.
Whether a source needs two Gaussians or not depends on its \hbb\ width.
Figure \ref{fig-hbwfp}{\it a} shows the probability $P$ that the
double-Gaussian model cannot improve the fit, as a function of FWHM(\hbb).
The 2435 sources below the dashed line ($\sim 32\%$ of the whole sample)
require two Gaussians at a significance greater than 3 $\sigma$.  For
sources with FWHM(\hbb) $\gtrsim$ 5000 \kms, one Gaussian is enough to
describe the profile of \hbb.  This is consistent with \citet{collin06}, who
find that the ratio FWHM/$\sigma$ ($\sigma$ is the line dispersion) of \hbb\
depends on its line width (see their Fig. 3). For sources with broader \hbb,
this ratio is close to 2.35, the value for a single Gaussian. As discussed
later, this means that broader sources have a weaker \hbi\ component.

We use Monte Carlo simulations to determine the detection threshold of \hbi.
We generate artificial \hb\ lines using two components [FWHM(\hbvb) is set to
2.5 FWHM(\hbi) as shown in Fig. \ref{fig-hbihbvb}], fit it using a single 
and a double Gaussian, and then calculate the probability $P$ as before. The 
EW ratio of \hbi/\hbvb\ is increased progressively until $1-P$ exceeds 
99.73\%. (All EW measurements refer to the continuum at 5100 \ang.) 
The critical EW ratio of \hbi/\hbvb\ depends on S/N, as shown in 
Figure \ref{fig-hbwfp}{\it b}. In order to detect an \hbi\ component with an 
EW that is 10\% larger than the EW of \hbvb, the S/N should be larger than 
$22.5$.  Thus, in the analysis below, we remove all sources with (1) 
S/N $< 22.5$, (2) \ks\ of double-Gaussian 20\% larger than the \ks\ of
Gauss-Hermite and $1-P < 99.73$\%, and (3) EW(Fe) $<$ 25 \ang\ (to insure
that the \feii\ measurements are as reliable as those in Paper I).
These three cuts left 1499, 811, and 568 sources, respectively. Note that
the fraction of sources that require a double-Gussian model (811/1499 $\approx$
54\%) is larger than the fraction when the S/N threshold is 10 (2435/7601 
$\approx$ 32\%), consistent with the simulations above: fainter \hbi\ can
be detected with higher S/N. The analysis below focuses on the most
stringent subset of 568 sources ($\sim 38\%$ of the sources with S/N
$>$ 22.5). The reduced \ks\ for the double-Gaussian model fit has a median
value of 1.004.

To constrain the relative strengths of \feii\ emission associated with the
very broad and intermediate-width components, we generated simulated spectra 
using two \feii\ components, and then fit them using only one component, whose 
width is fixed to that of the input intermediate-width component. The EW of 
the input very broad component is increased progressively until the reduced \ks\ 
of the fit exceeds 1 $\sigma$ from the expect value. For a typical \feii\ EW 
of 75 \ang\ and S/N $\approx$ 25 in our final sample, the flux ratio of \feii\ 
between the input very broad component and intermediate-width component is 
$\sim$0.3. This is the upper limit of the \feii\ emission coming from the 
very broad component. 

\begin{figure*}
  \includegraphics[angle=-90,width=0.98\textwidth]{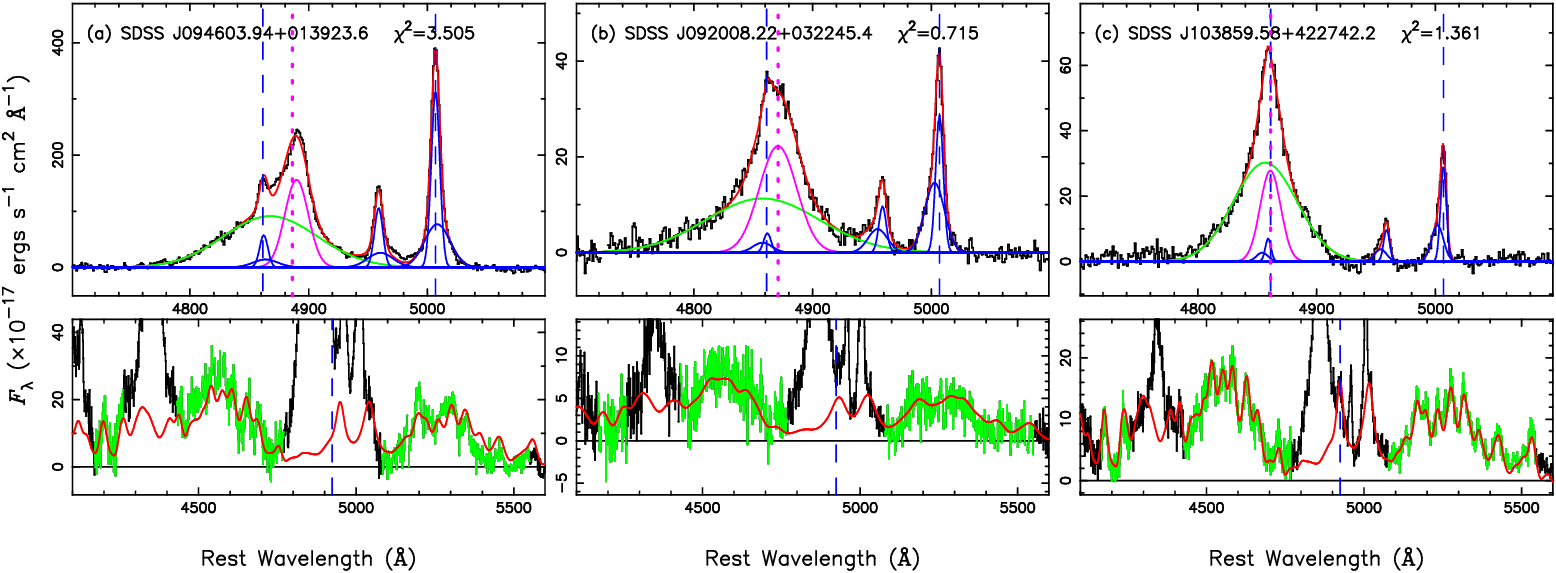}
  \caption{\footnotesize Examples of \feii\ measurement and emission-line
  fitting for ({\it a}) SDSS J094603.94$+$013923.6, ({\it b}) SDSS
  J092008.22$+$032245.4, and ({\it c}) SDSS J103859.58$+$422742.2.  For each
  source, the top panel shows multi-Gaussian fitting for \hb\ and \oiii.
  \hbvb\ is in green and \hbi\ is in magenta. The two blue dashed lines mark
  the rest-frame wavelength of \hb\ and \oiii\ $\lambda$5007. The magenta
  dotted line is the position of \hb\ at the same velocity as \feii.  Note the 
  consistency between the \hbi\ peak and the dotted line.
  \hbn\ and \oiii\ are in blue. The red line is the sum of each component. The
  bottom panel shows the emission-line spectrum after subtracting the power-law 
  continuum. The green portions of the spectrum denote the windows for fitting 
  the \feii\ emission, whose model is given in red.  The blue dashed line marks 
  the peak of \feii\ $\lambda$4924 at zero velocity shift. 
  }
  \label{fig-example}
\end{figure*}

Figure \ref{fig-example}
shows the emission-line fitting and \feii\ emission measurement for three
typical sources. Example {\it a} is an extreme case that has a large \feii\
redshift (1533$\pm$24 \kms) and \hb\ with an isolated red peak. The
\feii\ in example {\it b} is moderately redshifted (590$\pm$98 \kms), and its
\hb\ core shows only a strong red asymmetry rather than another peak. A more 
common situation is seen in example {\it c}, in which \feii\ and \hbi\ have no 
shift and the \hb\ profile is symmetrical.  Table \ref{tab-all} lists the 
measurements of the three sources shown in Figure \ref{fig-example}.

\begin{table*}
\begin{center}
  \caption{Line Measurements
  \label{tab-all}}
{\scriptsize
\begin{tabular}{cccccccccccccc}
  \hline\hline
 & &\multicolumn{3}{c}{\feii} & & \multicolumn{3}{c}{\hbi} & & 
\multicolumn{3}{c}{\hbvb} & \\
\cline{3-5} \cline{7-9} \cline{11-13}
SDSS Name & $z$ & EW & FWHM & Shift & & EW & FWHM & Shift & &
EW & FWHM & Shift & \ks \\
 & &(\ang)&(\kms)&(\kms)& &(\ang)&(\kms)&(\kms)& &(\ang)&(\kms)&(\kms)& \\ 
 \hline
094603.94$+$013923.6 & 0.220 & 60$\pm$1 & 1543$\pm$55 & 1533$\pm$24 & &
54$\pm$1 & 1428$\pm$18 & 1754$\pm$7 & & 
125$\pm$1 & 5730$\pm$ 58 & 400$\pm$25 & 3.506 \\
092008.22$+$032245.4 & 0.334 & 45$\pm$2 & 2548$\pm$251 & 590$\pm$98 & &
27$\pm$2 & 2281$\pm$103 & 589$\pm$42 & & 
42$\pm$2 & 6924$\pm$332 & $-$195$\pm$109 & 0.715 \\
103859.58$+$422742.2 & 0.221 & 72$\pm$1 & 1297$\pm$40 & $-$10$\pm$18 & &
13$\pm$1 & 1259$\pm$71 & $-$16$\pm$26 & &
44$\pm$1 & 3893$\pm$87 & $-$279$\pm$29 & 1.361 \\ \hline
\end{tabular}
}
\parbox{0.9\textwidth}{
\tablecomments{Table \ref{tab-all} is available in its entirety 
  electronically.}
}
\end{center}
\end{table*}

\section{Results and Discussions}

\begin{figure} 
  \includegraphics[angle=-90,width=0.47\textwidth]{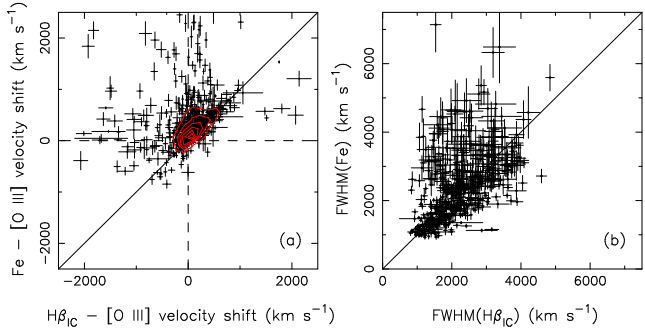}
  \caption{\footnotesize Correlations between ({\it a}) \feii\ and \hbi\
  shifts and ({\it b}) \feii\ and \hbi\ widths. The solid diagonal lines 
  denote that \feii\ and \hbi\ have the same shifts and widths. The contours
  in panel ({\it a}) show the density of the data points.} 
  \label{fig-hbife} 
\end{figure}

\begin{figure}
  \includegraphics[angle=-90,width=0.47\textwidth]{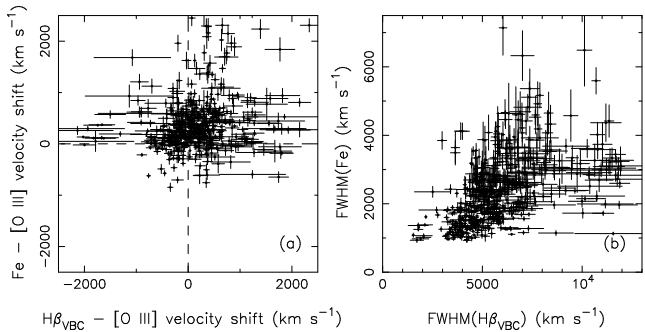}
  \caption{\footnotesize Plots of ({\it a}) \feii\ vs. \hbvb\ shifts and
  ({\it b}) \feii\ vs. \hbvb\ widths. The two lines have different shifts
  and widths.}
  \label{fig-hbvbfe}
\end{figure}

\begin{figure}
  \includegraphics[angle=-90,width=0.47\textwidth]{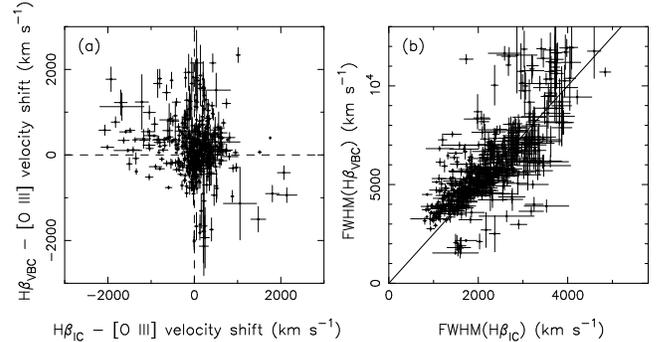}
  \caption{\footnotesize Plots of ({\it a}) \hbvb\ vs. \hbi\ shifts and
  ({\it b}) \hbvb\ vs. \hbi\ widths. The two components have different shifts
  and widths. The solid line in panel ({\it b}) denotes FWHM(\hbvb) $= 2.5$
  FWHM(\hbi).}
  \label{fig-hbihbvb}
\end{figure}

We find that \hbi\ and \feii\ emission have similar kinematics. The similarity 
can be seen not only in individual sources, but also statistically for the 
whole sample.  From examples {\it a} to {\it c} in Figure \ref{fig-example}, the
\hb\ profiles change progressively while the \feii\ shifts become lower and
lower. Figure \ref{fig-hbife}{\it a} shows a strong correlation between \hbi\
and \feii\ shifts. Pearson's correlation coefficient $r_{\rm P}$ is 0.22,
and the probability $P$ of a chance correlation is $< 1\times10^{-5}$. 
\hbi\ and \feii\ have approximately the
same shifts. The widths of \hbi\ and \feii\ are also well correlated and
roughly equal (Fig. \ref{fig-hbife}{\it b}); $r_{\rm P}$ = 0.48 and $P <
1\times10^{-5}$. Except for some sources to the upper left of the solid lines
whose errors are large, the majority of sources follow the relation that \hbi\ and
\feii\ have the same shifts and widths. By contrast, \feii\ and \hbvb\ have
different shifts and widths (Fig. \ref{fig-hbvbfe}). We also find
that the kinematic
connection between \hbi\ and \hbvb\ is complicated.  No correlation between
\hbi\ and \hbvb\ shifts is seen (Fig. \ref{fig-hbihbvb}{\it a}), but the
\hbi\ and \hbvb\ widths are strongly linked, such that FWHM(\hbvb) $\approx$
2.5 FWHM(\hbi) (Fig. \ref{fig-hbihbvb}{\it b}).

The observational results described above suggest a scenario in which the
conventional BLR consists of two components---an intermediate-line region 
(ILR) and a very broad-line region (VBLR).
If both regions are virialized, so that $R \propto v^{-2}$, then the ILR is 
about $2.5^2$ or 6.25 times farther from the center than the VBLR.  Because of 
its redshift, the kinematics of the ILR may be dominated by infall. \hb\ 
emission emerges from both the ILR and VBLR, while most of the \feii\ emission 
comes from the ILR.  

The ILR and VBLR defined in this paper are essentially similar to those
described in \citet{corbin95}, \citet{brotherton96}, \citet{sulentic00b},
and \citet{zhu08} but
differ from those in \citet{brotherton94} or \citet{sulentic99}, 
which refer to the \civ-emitting region.  Note that the ILR and VBLR of the 
\hb-emitting region are distinct from those of the \civ-emitting region
because the \civ\ ILR usually has the systemic redshift while the \civ\ VBLR 
shows a blueshift. 

\begin{figure}
  \includegraphics[angle=-90,width=0.47\textwidth]{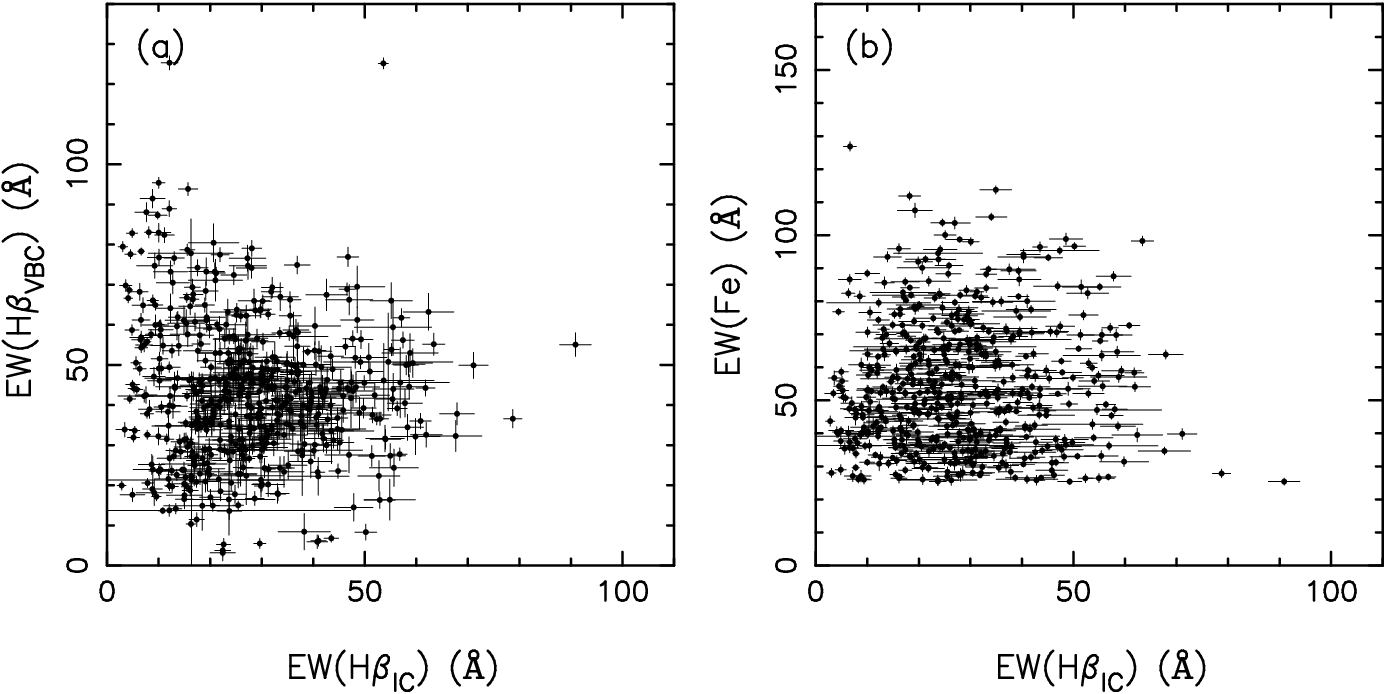}
  \caption{\footnotesize Plots of ({\it a}) \hbvb\ vs. \hbi\ EWs and
  ({\it b}) \feii\ vs. \hbi\ EWs. No correlations are found.}
  \label{fig-ew}
\end{figure}

The lack of correlation between the EWs of \hbi\ and \hbvb\ (Fig. 
\ref{fig-ew}{\it a}) strongly suggests that the two components are emitted 
from different regions. If both are photoionized, they must have different 
covering factors. The relative strength between \hbi\ and \hbvb\ determines 
the final \hb\ profiles.  In Figure \ref{fig-hbwfp}, sources broader than 
$\sim$5000 \kms\ can be fitted well using one Gaussian. This trend can be 
interpreted in our two-component BLR scenario. It is well known that sources 
with broad \hb\ tend to have weak \feii/\hb\ 
\citep[e.g.,][]{bg92,sulentic00a}.  The ILR is weak in these systems because
\feii\ is weak.  Their profiles show little deviation from a single Gaussian 
under the typical S/N level of SDSS spectra. The composite spectra of sources 
with large \feii\ redshifts, on the other hand, do show red asymmetry in the 
\hb\ profiles (see Fig. 13 of Paper I).  The variation in the relative 
strength of the \hbvb\ and \hbi\ components in different sources reflects
the competition between the two components,
although they apparently do so in such a manner that their kinematics
remained coupled.

It is of interest to note that the strengths of \hbi\ and \feii\
are not correlated (Fig. \ref{fig-ew}{\it b}). The wide range of 
\feii/\hbi\ ratios reflects either the complexity of the excitation mechanism
of \feii\ emission \citep[e.g.,][and references therein]{baldwin04} or 
large variations in quasar metallicities \citep[e.g.,][]{netzer07}.

\section{Summary}

We have studied the profiles of the \hb\ emission line using a large sample of
quasars selected from SDSS. Comparing the \hb\ profiles with the properties of 
\feii\ emission given in Paper I, we deduced the existence of two \hb\ 
emission regions---an intermediate-line region and a very broad-line region. 
The observational evidence can be summarized as follows:
\begin{enumerate}
\item{The velocity shifts and widths of the \hb\ intermediate-width component 
are approximately the same as those of \feii, indicating that they originate
from the same region.  However, the \feii/\hbi\ ratios vary greatly from 
object to object, reflecting variations in either metal abundance or 
excitation conditions in the \feii-emitting region.}
\item{The velocity width of the very broad component of \hb\ is roughly 
2.5 times larger than that of the intermediate-width component, but no 
correlation exists between their radial velocities.  The equivalent widths 
of the two components are also unrelated, suggesting that they have very 
different covering factors and geometry.  The conventional BLR seems to 
consist of two different, physically distinct regions.  We suggest 
that the intermediate-width component of \hb\ and \feii\ trace an infalling 
region in the outskirts of the BLR, likely located in between the molecular 
torus and the accretion disk.}
\end{enumerate}

The properties of the \hb-emitting region discussed here, in conjunction with 
those of \feii\ emission summarized in Paper I, offer important new 
constraints on models of the broad emission-line regions.  This will be 
the subject of a forthcoming paper.

\acknowledgments

We appreciate extensive discussions among the members of the IHEP AGN group. 
We thank an anonymous referee for helpful comments.
This research is supported by NSFC via NSFC-10325313, 10733010 and
10521001, and by CAS via KJCX2-YW-T03.

\end{document}